\documentclass[sigconf, authorversion,nonacm]{acmart}

\AtBeginDocument{%
  }

\usepackage{hyperref}
\usepackage{amsmath,amsfonts}
\usepackage{algorithmic}
\usepackage{graphicx}
\usepackage{textcomp}
\usepackage{xcolor}
\usepackage{svg}
\usepackage{subfiles}
\usepackage{multirow}
\usepackage{array}
\usepackage{soul}

\begin{document}

\title{Beyond Isolated Frames: Enhancing Sensor-Based Human Activity Recognition through Intra- and Inter-Frame Attention}

\author{Shuai Shao}
\affiliation{%
  \institution{University of Manchester}
  \city{Manchester}
  \country{UK}}
  \email{shuai.shao@manchester.ac.uk}

\author{Yu Guan}
\affiliation{%
  \institution{University of Warwick}
  \city{Coventry}
  \country{UK}}
\email{yu.guan@warwick.ac.uk}

\author{Victor Sanchez}
\affiliation{%
  \institution{University of Warwick}
  \city{Coventry}
  \country{UK}}
\email{V.F.Sanchez-Silva@warwick.ac.uk}


\begin{abstract}
  Human Activity Recognition (HAR) has become increasingly popular with ubiquitous computing, driven by the popularity of wearable sensors in fields like healthcare and sports. While Convolutional Neural Networks (ConvNets) have significantly contributed to HAR, they often adopt a frame-by-frame analysis, concentrating on individual frames and potentially overlooking the broader temporal dynamics inherent in human activities. To address this, we propose the intra- and inter-frame attention model. This model captures both the nuances within individual frames and the broader contextual relationships across multiple frames, offering a comprehensive perspective on sequential data. We further enrich the temporal understanding by proposing a novel time-sequential batch learning strategy. This learning strategy preserves the chronological sequence of time-series data within each batch, ensuring the continuity and integrity of temporal patterns in sensor-based HAR.
\end{abstract}




\maketitle

\section{Introduction}

In the field of ubiquitous computing, Human Activity Recognition (HAR) has become a core research, driven by advances in wearable and sensor technologies. These sensors, from smartwatches to advanced medical devices, are crucial in healthcare, sports training, and elderly care~\cite{Hammerla2015PDLearning, Majumder2017WearableMonitoring, Stavropoulos2020IoTReview}. Their influence has been instrumental in reshaping our approaches to monitoring, analyzing, and understanding human activities in real-world scenarios.

Convolutional Neural Networks (ConvNets) have become the mainstream in the field of sensor-based HAR~\cite{Wang2019DeepSurvey}, demonstrating a notable proficiency in feature extraction from sensor data. Their prowess in discerning intricate patterns in sequential data has revolutionized HAR, marking a significant leap over traditional  methods~\cite{Hammerla2016DeepWearables}. Despite these advances, HAR faces challenges, notably in segmenting continuous sensor data using the sliding window technique~\cite{Bulling2014ASensors}. This method, while popular, can segment activities that exceed the fixed frame lengths, potentially losing vital transitional and contextual data~\cite{Dehghani2019ASensors}. The sliding window size dilemma further complicates this segmentation challenge, as smaller windows may miss complete activities, while larger ones could mix unrelated activities. Recent research emphasizes the need for models that capture both detailed and broad activity sequences due to this segmentation issue~\cite{Hiremath2020OnRecognition, DasAntar2019ChallengesReview}.

\begin{figure}[tbp]
\centerline{\includegraphics[width=0.45\textwidth]{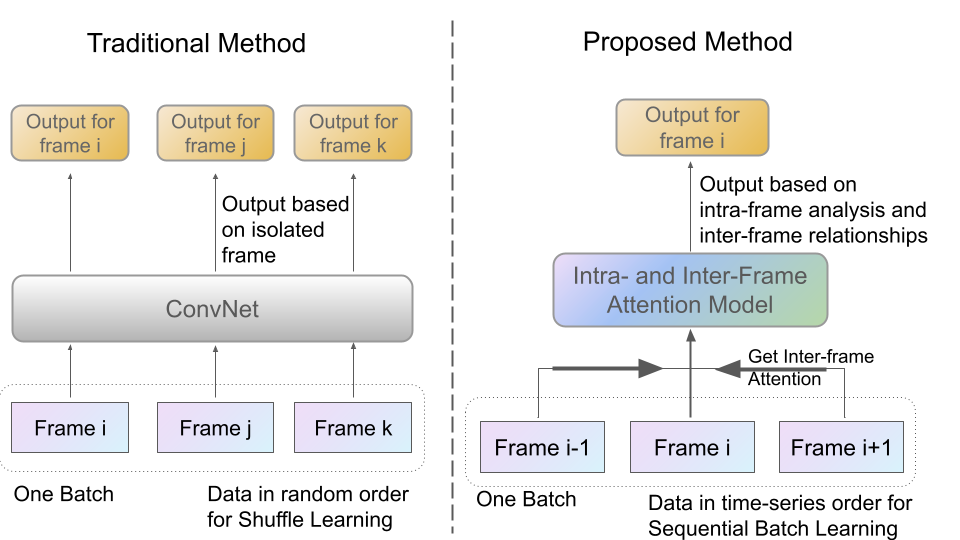}}
\caption{Comparative overview of the traditional method vs. our proposed method.}
\label{method_compare}
\end{figure}

This paper introduces a novel intra- and inter-frame attention model designed to capture the subtle nuances of each frame and their collective dynamics within a batch. By implementing a time-sequential batch learning strategy, our method preserves the temporal sequence of frames, which is crucial for detecting subtle temporal patterns during training. As illustrated in ~\autoref{method_compare}, our approach contrasts traditional deep learning methods that typically generate outputs based on isolated frames. Our model uniquely considers both intra- and inter-frame relationships, enhancing the training process. Further refinements include the incorporation of a combined loss function, which is designed to boost the robustness and accuracy of the model.

The main contributions of our proposed method are summarized below:
\begin{enumerate}
    \item We propose and design the intra- and inter-frame attention model, capturing details within and between frames within a batch.
    \item We introduce a time-sequential batch learning strategy, which ensures the chronological order of frames within a batch, preserving essential temporal information.
    \item A combined loss function to improve the training process, enhancing the robustness and accuracy of HAR.
    \item Validation of our method through comprehensive empirical testing and an ablation study to highlight the importance of each model component.
\end{enumerate}

\section{Related Work}

ConvNets have been transformative in sensor-based HAR, where they transitioned from image processing to adeptly handling time-series data feature extraction~\cite{Yang2015DeepRecognition., Shao2023ConvBoost:Recognition, Aguileta2019Multi-SensorSurvey}. While they reduce the need for manual feature engineering and enhance multi-sensor data interpretation, they struggle with capturing long-term dependencies and optimal frame sizing, which can affect recognition accuracy and lead to potential overfitting.

Attention mechanisms have revolutionized HAR by dynamically prioritizing different segments of input data based on their contextual relevance. For instance, AttnSense~\cite{Ma2019AttnSense:Recognition} integrates attention with ConvNets and GRUs, effectively capturing both spatial and temporal dependencies. Despite their strengths, many attention-based models focus primarily on isolated frame analysis and may overlook extensive temporal patterns that span multiple frames.

While recent advancements have highlighted the benefits of advanced batch training strategies, the application to frame-based ConvNets models remains limited. Pellatt and Roggen~\cite{Pellatt2020CausalBatch:Recognition} introduced 'CausalBatch', a training method that significantly enhances the performance of LSTM-based networks by structuring batches to maintain temporal continuity. Moreover, the 'BatchFormer' module introduced by Hou et al.~\cite{Hou2022BatchFormer:Learning} offers a compelling direction for ConvNets through its application in computer vision. BatchFormer utilizes transformer technology to explore and utilize sample relationships within each mini-batch, enriching the representation learning process. Although originally applied in the context of visual data, this method inspires potential adaptations for sensor-based HAR, where similar challenges in data scarcity and the need for robust feature extraction prevail.

\section{Methodology}
Although the sliding window approach~\cite{Bulling2014ASensors} is commonly used in HAR, it often fails to capture activities spanning multiple frames, losing crucial interconnections and long-range contextual information.

In response, our primary contribution is the intra- and inter-frame attention model, which takes advantage of time-sequential batch learning to overcome the constraints of individual frames. This model offers a detailed analysis of both the nuances within frames and the broader relationships across them.

\subsection{Intra- and Inter-Frame Attention Model}

\begin{figure}[tbp]
\centerline{\includegraphics[width=0.5\textwidth]{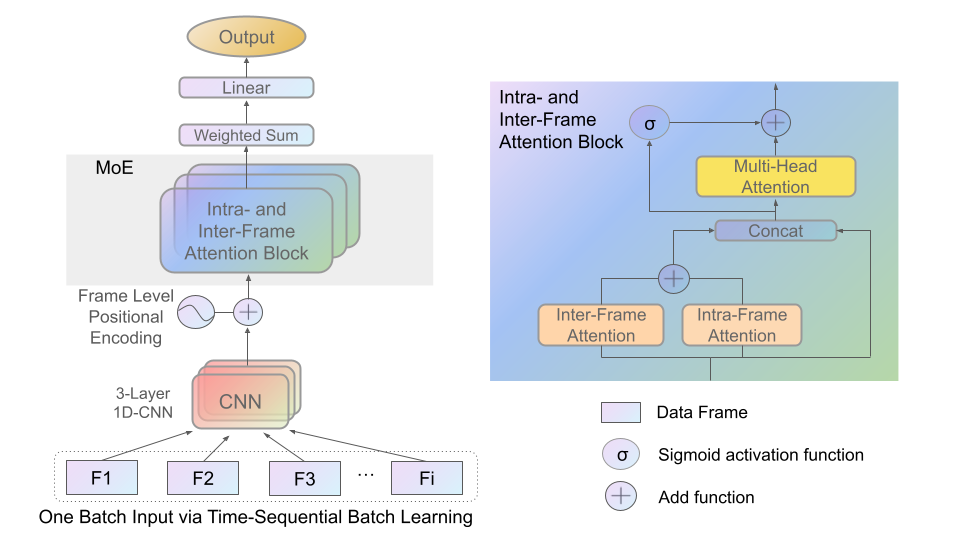}}
\caption{An overview of our proposed Intra- and Inter-Frame Attention Model.}
\label{model}
\end{figure}

Traditional HAR methods often analyze activities as isolated events, possibly leading to fragmented insights. Our proposed model, however, focuses on the continuous context of activities, aiming for a more comprehensive understanding. An in-depth description of our methodology is provided in~\autoref{model}. This model integrates positional encoding, intra- and inter-frame attention, and the Mixture of Experts (MoE), each contributing to the model's effectiveness in recognizing complex activity patterns. 

\subsubsection{Positional Encoding at Frame Level}

Positional encoding is crucial for integrating sequence order information into the model. Unlike the traditional within-frame encoding, we introduce positional encoding at the frame level in our approach. This ensures that each frame within a batch is endowed with a unique positional representation, allowing the model to discern not only the content of each frame but also its relative position in the sequence.

\subsubsection{Attention Mechanisms: Bridging Intra- and Inter-Frame Dynamics}

Inspired by the success of self-attention applications~\cite{Vaswani2017AttentionNeed, Murahari2018OnRecognition, Abedin2020AttendSensors}, we recognize the potential of attention mechanisms to uncover dependencies within time-series data. Based on this insight, we develop the intra- and inter-frame attention block. Our model utilizes intra-frame attention to focus on details within individual frames and inter-frame attention to explore dependencies across multiple frames. This dual attention strategy ensures a comprehensive understanding of activities, essential for effective HAR.

Intra-frame attention examines individual data points within a frame, using a matrix representation \( {X} \) of the data in a frame to compute attention scores:
\begin{equation}
A_{\text{intra}} = \text{softmax}(W_2 \tanh(W_1 {X} + b_1) + b_2)
\end{equation}
Here, \( W_1 \) and \( W_2 \) are weight matrices, and \( b_1 \) and \( b_2 \) are bias terms, which are parameters learned during training.

Inter-frame attention, meanwhile, assesses relationships between frames using a scaled dot-product mechanism:

\begin{equation}
A_{\text{inter}} = \text{softmax}(\frac{Q K^T}{\sqrt{d}})V
\label{attention}
\end{equation}

\noindent where Query (Q), Key (K), and Value (V) are matrices derived from the frame data within a batch, \( d \) is the dimension of the embedding.

To capture the full spectrum of sensor data relations, we blend insights from both intra- and inter-frame dynamics:

\begin{equation}
A_{\text{com}} = \alpha \times A_{\text{inter}} + (1 - \alpha) \times A_{\text{intra}},
\end{equation}

\noindent where $\alpha$ is a trainable parameter balancing the two forms of attention.

This combined attention feeds into a multi-head attention mechanism, enhancing the representation of each frame within the context of its batch:

\begin{equation}
X_{\text{att}} = Concate(\Bar{X}, A_{\text{com}}).
\end{equation}

\begin{equation}
A_{\text{mul}} = MulAtt(X_{\text{att}}, X_{\text{att}}, X_{\text{att}}),
\end{equation}
where the $MulAtt$ denotes the aggregation of multiple attention heads.

A gating mechanism then adjusts the influence of multi-head attention based on the temporal characteristics of the data, effectively merging the information:

\begin{equation}
G = \sigma(W_g X_{\text{att}} + b_g),
\end{equation}
where $W_g$ is the gating weight matrix, $b_g$ is the bias, and $\sigma$ is the sigmoid activation function. This gating score, $G$, indicates the proportion of influence the multi-head attention has on the model's output. 

The final output of the proposed attention block, $O_{\text{gated}}$, is formulated as:
\begin{equation}
O_{\text{gated}} = G \odot A_{\text{mul}} + (1 - G) \odot X_{\text{enhanced}}.
\end{equation}

By enhancing the temporal and contextual understanding of HAR data, this model provides a more robust framework for analyzing complex human activities.

\subsubsection{Amplifying Frame Relations via Mixture of Experts}

To tackle the complexity of human activities that may involve multiple classes within the same batch, our model incorporates a Mixture of Experts (MoE). This approach allows for diverse analytical perspectives, enhancing the model's capability to recognize intricate patterns across varied activities.

The MoE extends our core intra- and inter-frame attention mechanism by adding specialized interpretations for each unique pattern identified in the data. Each expert processes the data independently and outputs a distinct result, donated as $e_{i}$. A gating mechanism furnishes a weight set, \(W = \{w_1, w_2, \ldots, w_n\}\), reflecting the pertinence of each expert's interpretation. The result can then be expressed as: $O_{\text{MoE}} = \sum_{i=1}^{n} w_i \cdot e_i.$, where $O_{\text{MoE}}$ represents the aggregated insights from all experts, providing a comprehensive view of the activity being analyzed. This method ensures that the model can adapt to and effectively analyze complex scenarios with multiple activity types present simultaneously.

\subsection{Time-Sequential Batch Learning}
Mainstream sensor-based HAR training techniques often rely on random frame selection during training to mitigate overfitting~\cite{Ordonez2016DeepRecognition, Hammerla2016DeepWearables, Murahari2018OnRecognition, Abedin2020AttendSensors, Shao2023ConvBoost:Recognition}. While effective in certain scenarios, this approach can disrupt the inherent temporal sequences present in activity data, potentially affecting the model's ability to recognize sequential patterns.

\begin{figure}[tbp]
\centerline{\includegraphics[width=0.5\textwidth]{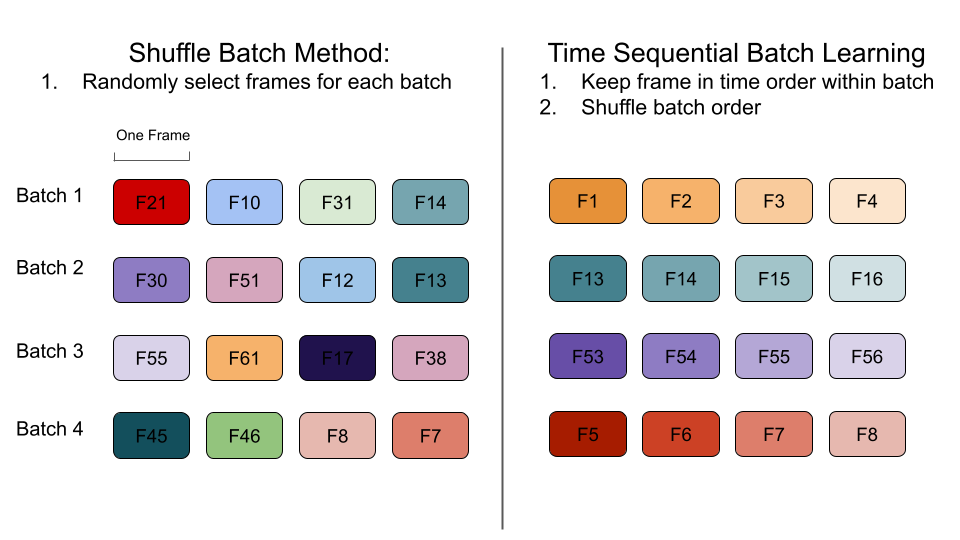}}
\caption{Comparison of Shuffle Learning vs. Time-Sequential Batch Learning (Varying shades of colour indicate the progression of time, best view in colour).}
\label{fig:seq_vs_shuffle}
\end{figure}

Recognizing the significance of sequential data in HAR, we propose Time-Sequential Batch Learning. This training strategy prioritizes the chronological integrity of time-series data, ensuring that frames within a batch are processed in their temporal sequence. This approach is pivotal in preserving the continuity and richness of sequential data.

To strike a balance between maintaining temporal sequences and preventing overfitting, we introduce a randomized batch selection strategy. While the order of frames within a batch remains chronological, the sequence of these batches is randomized for each training epoch. Our proposed approach seeks to combine the advantages of preserving time-series sequences with the benefits of randomization, to ensure that time-series details are effectively captured without overfitting. As illustrated in~\autoref{fig:seq_vs_shuffle}, the Time-Sequential Batch Learning maintains the chronological order of frames within each batch during a model training phase, in contrast to the random frame selection in traditional Shuffle Learning.

\subsection{Combined Loss}

In sensor-based HAR, the distribution of activity types in datasets can be uneven, with some activities being underrepresented. This imbalance can lead to biased learning, where the model overly focuses on the majority class and fails to consider less frequent activities.

To address this issue, we utilize the Focal Loss~\cite{Lin2017FocalDetection}, which modifies the standard Cross-Entropy loss (CE) to emphasize harder, often misclassified examples. The Focal Loss formula is: $FL(p_t) = -\beta (1 - p_t)^\gamma \log(p_t)$, where \( p_t \) represents the model's predicted probability for the actual class, \( \beta \) scales the importance of negative examples, and \( \gamma \) increases the focus on difficult examples. We combine the Focal Loss with the Cross-Entropy loss to create a balanced loss function: $L_{com} = (1 - \lambda) \times CE + \lambda \times FL$, where \( \lambda \) is a tunable parameter that balances the two loss types. This combined approach aims to improve model robustness and accuracy across a varied range of activities, ensuring fair treatment of all classes regardless of their frequency.

\section{Experiment}
\subsection{Datasets}

In our experiments, we employ four public datasets: Opportunity (OPP)~\cite{Chavarriaga2013TheRecognition}, Growing Old Together Validation (GOTOV)~\cite{Paraschiakos2020ActivityElderly}, Hospital~\cite{Yao2017EfficientNetworks}, and Physical Activity Monitoring Dataset (PAMAP2)~\cite{Reiss2012IntroducingMonitoring}. Each of these datasets corresponds to a unique HAR application and offers a varied set of challenges that help validate our method and compare it with the existing state-of-the-art. 

Opportunity (OPP) is recognised as one of the more challenging wearable-based HAR datasets, OPP exhibits pronounced imbalances in class distributions. Adhering to the methodologies outlined in~\cite{Hammerla2016DeepWearables, Guan2017EnsemblesWearables}, we employ a hold-out evaluation following the same settings. Growing Old Together Validation (GOTOV) focuses on daily activities from elderly participants, capturing 16 distinct activities across thirty-five subjects. In our experiment, six participants, lacking complete sensor data, are omitted. Consequently, we utilize the data from twenty-nine participants. We follow the same hold-out settings in~\cite{Shao2023ConvBoost:Recognition}. Hospital dataset is integral to care applications as it contains activity data from 12 hospitalized elderly patients. They were equipped with inertial sensors, and each performed 7 distinctive activities. We follow the same settings as in~\cite{Yao2017EfficientNetworks}, we use data from the initial 8 participants for training and the subsequent 3 for testing. The remaining data are set aside for validation purposes. Physical Activity Monitoring Dataset (PAMAP2) is a widely used wearable-based HAR dataset, which covers 12 daily activities such as running, walking, lying, and sitting, gathered from nine subjects. As in the methodologies of ~\cite{Hammerla2016DeepWearables, Guan2017EnsemblesWearables}, we apply the same hold-out evaluation approach.

\subsection{Evaluation Metric}

In evaluating the effectiveness of our proposed approach across all conducted experiments, we predominantly rely on the mean F1 score as the central performance metric. The mean F1 score serves as a balanced measure, capturing both precision and recall, and is especially crucial when there's an uneven class distribution or when false negatives and false positives have differing impacts. It is mathematically represented as: $\bar{F}_1 = \frac{1}{C} \sum_{c=1}^{C} \frac{2TP_c}{2TP_c + FP_c + FN_c}$, where \( C \) represents the total number of activity classes. For each specific class \( c \), \( TP_c \), \( FP_c \), and \( FN_c \) denote the counts of true positive, false positive, and false negative predictions, respectively. Using this metric ensures a comprehensive understanding of our model's capacity to correctly identify and distinguish between different human activities.

\subsection{Implementation Details}

In our experimental setup, we train our model end-to-end for 150 epochs using mini-batches of size 128 and the AdamW optimizer~\cite{Loshchilov2017DecoupledRegularization}. We initialize the learning rate to $10^{-3}$, and apply the ReduceLROnPlateau scheduling strategy from PyTorch, which halves the learning rate if there's no improvement in loss for 10 epochs. Our model uses a feature map size of 128, deploys 8 multi-head attention heads, and utilizes 8 experts in the MoE layer, with a dropout rate 0.5. The combined loss weighting coefficient \( \lambda \) varies by dataset: 0.5 for OPP, 0.2 for GOTOV, 0.3 for Hospital, and 0.1 for PAMAP2. The focal loss parameters \( \alpha \) and \( \gamma \) are set to 0.25 and 2, respectively. Data preprocessing involves normalizing to zero mean and unit variance. Data segmentation into frames uses a sliding window approach, with a 50\% overlap for OPP, GOTOV, and Hospital. Specifically, for the OPP dataset, we follow~\cite{Murahari2018OnRecognition, Abedin2020AttendSensors}, the window size is 24 samples. Both the GOTOV and Hospital datasets use a window size equivalent to 1 second, resulting in sizes of 84 and 20 samples, respectively. In the case of the PAMAP2 dataset, we follow~\cite{Hammerla2016DeepWearables}, employing non-overlapping sliding windows of 5.12 seconds duration and maintaining a one-second step between adjacent windows, which translates to a 78\% overlap.

\subsection{Model Comparison}

\begin{table}[htbp]
\centering
\caption{Mean F1 results of different models on various datasets}
\begin{tabular}{|p{2.5cm}|c|c|c|c|}
\hline
\textbf{Model} & \textbf{OPP} & \textbf{GOTOV} & \textbf{HOSPITAL} & \textbf{PAMAP2} \\ 
\hline
CNN~\cite{Yang2015DeepRecognition.} & 62.08 & 75.32 & 63.54 & 81.05 \\ 
\hline
ConvLSTM~\cite{Ordonez2016DeepRecognition} & 63.12 & 72.49 & 63.92 & 79.04 \\ 
\hline
Att. Model~\cite{Murahari2018OnRecognition} & 64.88 & 73.58 & 64.51 & \textbf{88.46} \\ 
\hline
Transformer & 61.05	& 73.22 & 63.85 & 83.26 \\
\hline
AD(CIE + AGE)~\cite{Abedin2020AttendSensors} & 65.82 & 76.62 & 65.07 & 87.62 \\ 
\hline
AD(CIE + AGE + CenterLoss)~\cite{Abedin2020AttendSensors} & 65.77 & 76.05 & 65.37 & 87.51
 \\ 
\hline
Ours & \textbf{69.21} & \textbf{86.15} & \textbf{66.53} & 85.13 \\ 
\hline
\end{tabular}
\label{result_table}
\end{table}

In our evaluation, we ensured fairness by re-implementing models from their public GitHub repositories and adapting them to our settings using the PyTorch library. For instance, the Transformer model is directly implemented in PyTorch, while the AD(CIE+AGE) model, including the addition of center loss in the AD (CIE+AGE+CenterLoss) model, is adapted from existing frameworks. We refrained from using external techniques such as data augmentation to focus solely on the inherent capabilities of each model.

From the results presented in ~\autoref{result_table}, we can gain insights into how various HAR models perform across multiple datasets. Traditional ConvNet models, such as CNN~\cite{Yang2015DeepRecognition.} and ConvLSTM~\cite{Ordonez2016DeepRecognition}, primarily designed for individual frame analysis, inherently lack the capability to capture intricate inter-frame relationships. This limitation is evident in datasets like OPP, GOTOV, and Hospital. On the OPP dataset, our method achieves a mean F1 score of 69.21\%, a notable improvement over CNN's 62.08\% and ConvLSTM's 63.12\%. Similarly, on the GOTOV dataset, our model achieves an F1 score of 86.15\%, surpassing CNN's 75.32\%. The trend is consistent on the Hospital dataset, where our model's F1 score of 66.53\% contrasts against CNN's 63.54\% and ConvLSTM's 63.92\%.

Furthermore, our approach still stands out when compared to advanced models. The Att. Model~\cite{Murahari2018OnRecognition} and AD (CIE+AGE)/AD (CIE+AGE+CenterLoss)~\cite{Abedin2020AttendSensors}, despite their advancements, still rely on frame-by-frame methods, which limits their ability to capture broader temporal patterns. On the OPP dataset, the Att. Model obtains a mean F1 score of 64.88\%, and 73.58\% on the GOTOV dataset, which are 4.33\% and 12.57\% lower than our results, respectively. The Transformer model, while transformative in many NLP tasks, exhibits only slight differences from conventional ConvNets in HAR. On datasets like Hospital, its performance is competitive, but it trails on OPP and GOTOV, scoring 61.05\% and 73.22\%, respectively. In comparison to state-of-the-art models like AD (CIE+AGE) and AD (CIE+AGE+CenterLoss), our approach demonstrates superior performance on datasets such as OPP and GOTOV. Specifically, our model achieves scores of 69.21\% on OPP and 86.15\% on GOTOV, outstripping AD (CIE+AGE)'s scores of 65.82\% and 76.62\% and AD (CIE+AGE+CenterLoss)'s scores of 65.77\% and 76.05\% on the respective datasets. These outcomes underscore the strength of our model, highlighting its unique capability to harness both intra- and inter-frame relationships, and setting it apart from other recent models across various datasets.

\begin{figure}[tbp]
\centering
\includegraphics[width=\linewidth]{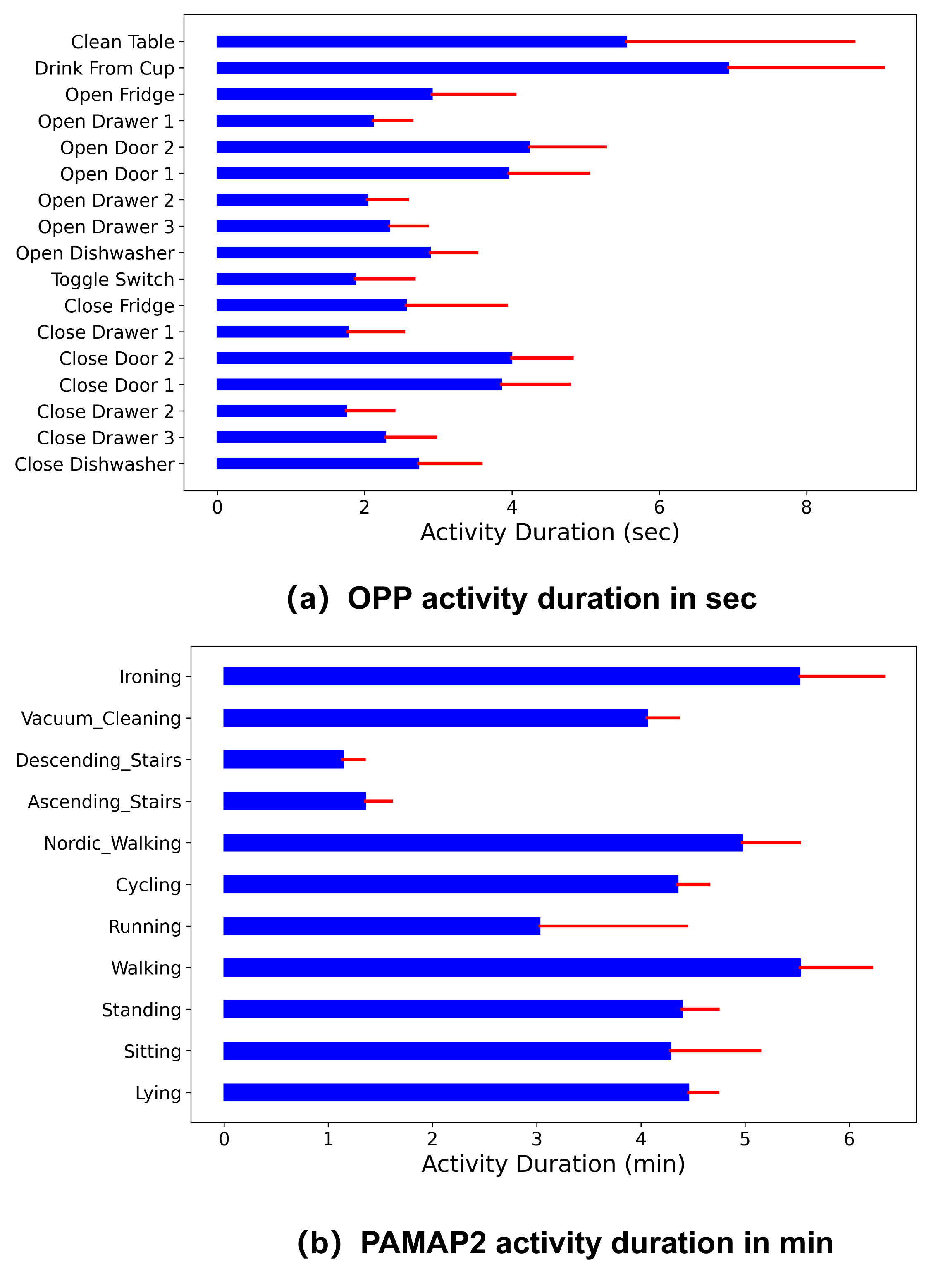}
\caption{The overview of the mean duration of each activity from OPP and PAMAP datasets, complemented by standard deviations, underscoring the central tendency and variability of activity duration. Here, OPP duration is expressed in seconds, while PAMAP2 duration is in minutes.}
\label{duration}
\end{figure}

\begin{figure*}[tbp]
\centering
\includegraphics[width=\linewidth]{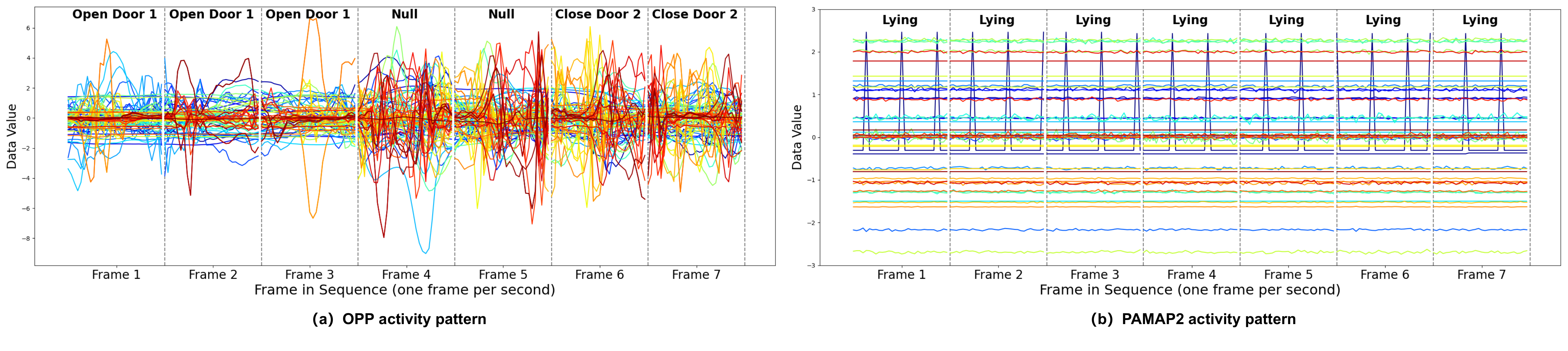}
\caption{The overview of the activity patterns from OPP and PAMAP datasets, emphasizing the temporal details and data characteristics, with unique colours denoting different data dimensions.}
\label{act_pattern}
\end{figure*}

While our model has shown strong performance across various benchmarks, its effectiveness is somewhat limited on datasets like PAMAP2, which predominantly consists of prolonged, repetitive activity patterns. As illustrated in subplots (a) and (b) in ~\autoref{duration}, the activities in the OPP dataset, represented in seconds, align well with our model’s strengths in capturing complex temporal dynamics and non-repetitive sequences. In contrast, the activities in PAMAP2, marked in minutes, involve extended periods of repetitive motions such as walking or cycling. Delving deeper into subplots (a) and (b) in ~\autoref{act_pattern}, the OPP dataset exhibits nuanced inter-frame variations, emphasizing the intricacies and complexities inherent in its data. These variations underscore the need for a model capable of capturing such fleeting dynamics. In contrast, the PAMAP2 dataset predominantly features consistent, recurring patterns, suggesting a different set of challenges where recognizing long-standing repetitive activities becomes paramount.

Overall, while the current ConvNets have set the foundation, and newer models have built upon this, our approach introduces a significant advancement by emphasizing the crucial role of intra- and inter-frame dynamics, yielding promising results in handling complex datasets and enriching the ongoing advancements in HAR.

\subsection{Ablation Studies}

\begin{figure}[tbp]
\centering
\includegraphics[width=\linewidth]{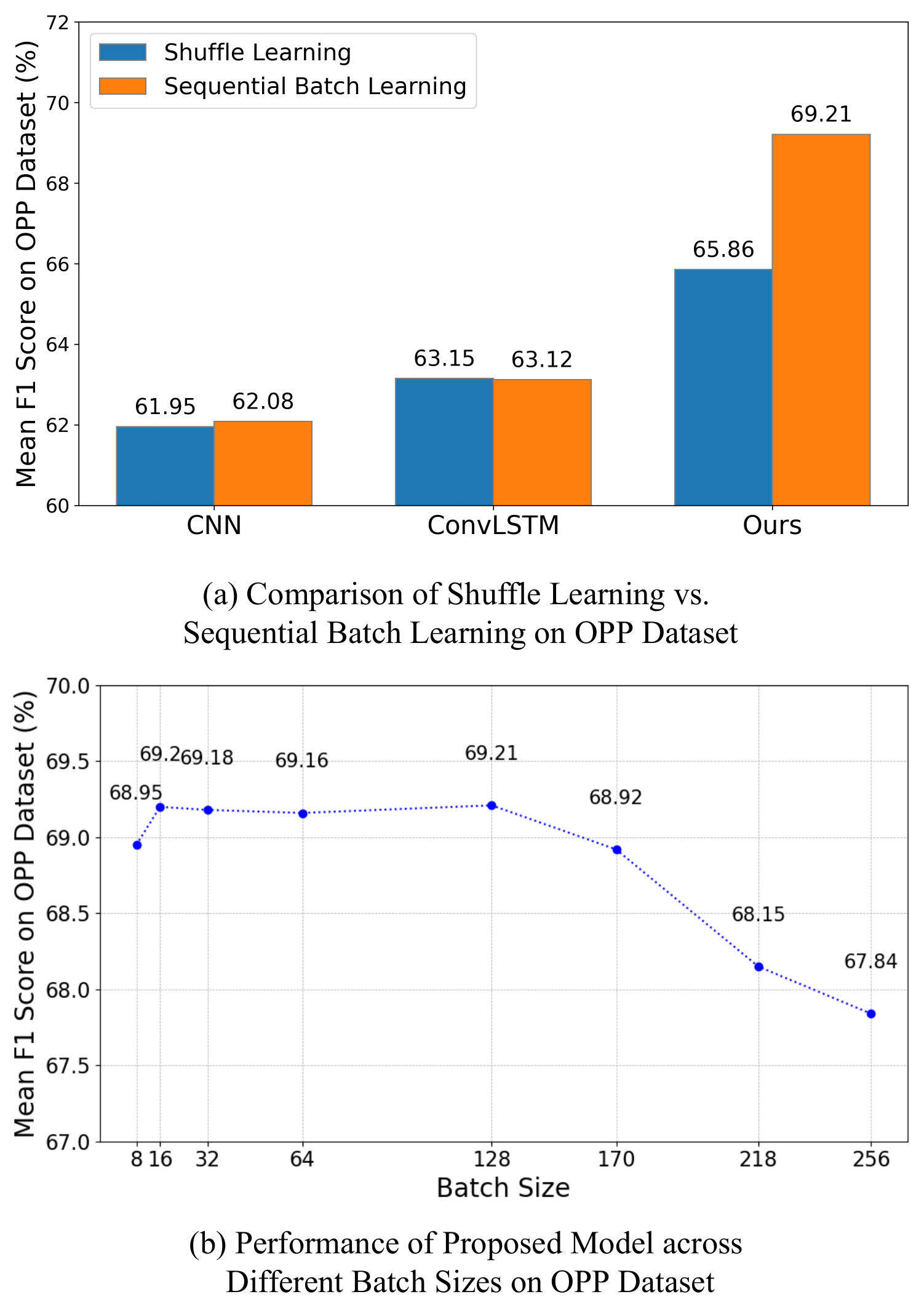}
\caption{The result of our ablation study.}
\label{fig:ablation_study}
\end{figure}

\subsubsection{Time Sequential Batch Learning Study}

In many deep learning approaches, shuffling data frames during training is a conventional protocol, typically to mitigate overfitting and boost generalization. However, our model relies on understanding temporal relationships within sequentially ordered batches, making the frame order critical for its performance.

As illustrated in \autoref{fig:ablation_study} (a), traditional ConvNets perform similarly under both shuffled and time-sequential learning, indicating their performance is not significantly affected by the order of frames. Conversely, our model benefits markedly from time-sequential batch learning, showing a notable increase of 3.35\% in mean F1 score on the OPP dataset compared to shuffled learning.

This improvement highlights our model's ability to capture intra- and inter-frame dynamics more effectively when trained with time-sequential batches. Adopting this strategy helps our model represent real-life temporal dynamics more accurately and reduces the risk of overfitting, proving essential for models that rely on temporal order.

\subsubsection{Impact of Batch Size on Model Performance}

The influence of batch size on the performance of our model is depicted in ~\autoref{fig:ablation_study} (b). Our model, grounded in the intra- and inter-frame attention mechanisms, heavily relies on sequential frames within a batch to discern meaningful relationships. This dependency is evident from the optimal performance observed for batch sizes ranging from 16 to 128.

However, as the batch size increases, especially beyond 170, there is a noticeable decline in recognition accuracy. For datasets like OPP, characterized by sporadic activities, a larger batch size tends to include frames from various activity classes within the same batch. If a batch contains various activity classes, the distinct temporal dynamics that the model aims to capture could be misrepresented, which may lead to inaccuracy when identifying relationships between frames.

\subsubsection{Component-wise Analysis}

\begin{table}[tbp]
\centering
\caption{Component-wise Ablation Results on the OPP Dataset}
\begin{tabular}{|p{6cm}|l|}
\hline
HAR   Models                                                  & $\bar{F}_1$    \\ \hline
Baseline ( CNN + MulAtt.)                                & 63.15 \\ \hline
Ours (Intra-Frame Att.)                                  & 64.18 \\ \hline
Ours (Inter-Frame Att.)                                  & 66.23 \\ \hline
Ours (Intra- and Inter-Frame Att.)                       & 67.05 \\ \hline
Ours (ALL) & 69.21 \\ \hline
\end{tabular}
\label{ablation_table}
\end{table}

~\autoref{ablation_table} provides a comprehensive component-wise ablation study of our model on the OPP dataset, highlighting the individual and collective contributions of the various components. Utilizing our model's overview graph in~\autoref{model} as a reference, we establish the ConvNet (CNN) combined with Multihead Attention as our baseline for isolated frame analysis. It is evident that the introduction of intra-frame attention offers a performance improvement compared to this baseline method. 

With the integration of inter-frame attention, there is a noticeable improvement in model performance. By adding frame-level positional information, our model emphasizes the importance of understanding the temporal frames within the batch. This ensures that the frames are analyzed not just as isolated instances, but in relation to their surrounding frames.

When both intra- and inter-frame attentions are combined, the model achieves a mean F1 score of 67.05\%, marking a 3.9\% improvement over the baseline. This highlights the strength of combining these attention mechanisms and underscores our primary contribution: the integration of intra- and inter-frame dynamics for a comprehensive understanding of human activities.

While our primary focus revolves around the intra- and inter-frame attention mechanisms, the complementary components further refine the model's efficacy. The integration of the MoE with attention mechanisms results in a significant performance boost. MoE operates by assigning different experts to specialize in various data subspaces. Each expert can provide a unique perspective or view on the data, ensuring that even within larger batches with diverse activities, the model can capture the nuances effectively. Lastly, the introduction of the combined loss, when paired with the other components, achieves the pinnacle of performance, indicating its role in further refining the model's training dynamics.

\section{Discussion and Conclusion}
The field of sensor-based HAR has relied heavily on ConvNets, focusing primarily on individual frame-by-frame analyses. Although this method has its strengths, it tends to overlook the broader, interconnected temporal dynamics that provide context across different activities. Our proposed model shifts from this traditional approach by focusing on both intra- and inter-frame attention. It captures the long-range contextual information that flows through the data, linking frames together in a batch to present activities as seamless sequences rather than isolated moments. This approach not only offers a clearer and more accurate understanding of activities but also taps into the subtle temporal patterns more effectively.

Our proposed model offers notable advantages, but also faces challenges. A primary concern is the architectural complexity that focuses on detecting inter-frame relationships, which can escalate both computational demands and training time. Furthermore, our model is designed to thrive when frames within a batch display a diverse range of patterns. However, if the frames tend to be repetitive or lack variation over extended periods, as illustrated in ~\autoref{act_pattern} (b), the model's performance may not reach its full potential.


In conclusion, our research introduces a fresh and comprehensive perspective to HAR, emphasizing the importance of long-range contextual information. The intra- and inter-frame attention model stands out as a significant advancement, enriching the HAR methodology and setting the stage for future explorations. This approach doesn't just enhance our understanding of human activities; it also opens up new possibilities for making activity recognition more holistic and insightful.


\balance
\bibliographystyle{ACM-Reference-Format}
\bibliography{references}

\end{document}